\documentclass[prb,aps,preprint,showpacs]{revtex4}
\usepackage{graphicx}
\usepackage{textcomp}

\usepackage{graphicx}
\usepackage{dcolumn}
\usepackage{bm}
\usepackage{color}
\begin{document}
\title{Structural and electronic properties of hybrid graphene and boron nitride nanostructures on Cu}

\author{Yan Li$^1$}
\author{Riccardo Mazzarello$^{1,2}$}\email{mazzarello@physik.rwth-aachen.de}

\affiliation{$^1$ Institute for Theoretical Solid State Physics, RWTH Aachen University, D-52074 Aachen, Germany\\
$^2$ JARA-FIT and JARA-HPC, RWTH Aachen University, D-52074 Aachen, Germany}

\date{\today}

\begin{abstract}

Recently, two-dimensional nanostructures consisting of alternating graphene and boron nitride (BN) domains have been synthesized.
These systems possess interesting electronic and mechanical properties, with potential applications in electronics and 
optical devices. 
Here, we perform a first-principles investigation of models of BN-C hybrid monolayers 
and nanoribbons deposited on the Cu$(111)$ surface, a substrate used for their growth in said experiments.
For the sake of comparison, we also consider BN and BC$_2$N nanostructures.
We show that BN and BC$_2$N monolayers bind weakly to Cu$(111)$, whereas monolayers 
with alternating domains interact strongly with the substrate at the B-C interface, due
to the presence of localized interface states. 
This binding leads to a deformation of the monolayers and sizable n-doping.
Nanoribbons exhibit a similar behaviour. Furthermore, they also interact significantly with the substrate at the edge,
even in the case of passivated edges. 
These findings suggest a route to tune the band gap and doping level of BN-C hybrid models based on the interplay between
nanostructuring and substrate-induced effects.

\end{abstract}
\maketitle

\section{Introduction}

Novel two-dimensional materials exhibiting promising properties for applications in information technology 
have been synthesized recently. 
Among them, graphene~\cite{Novoselov1} stands out for its versatility 
and its remarkable electronic and transport properties~\cite{CastroNeto,Morozov}.
Perfect monolayer (ML) graphene has a gapless spectrum prohibiting standard transistor applications.
A band gap can be induced by doping and/or by nanostructuring,
e.g. by fabricating one-dimensional nanoribbons (NRs)~\cite{Son,Jiao,Cai,Li,Ritter}. 
Hexagonal boron nitride (BN) MLs have been studied intensively too: they display exceptional
chemical stability and insulating behavior, with a large band gap of 5.7 eV.
It is also possible to tune the band gap of BN by making NRs with nanometer-sized width.

Recently, hexagonal BN-graphene hybrid structures, consisting of domains of BN and graphene, have been
synthesized experimentally~\cite{Ci,Liu}. These structures exhibit interesting electronic and mechanical
properties, which could be exploited in novel, flexible electronic and optical devices.
Furthermore, the properties of these systems can be tailored by changing the relative content      
of BN and graphene. We refer to these hybrid systems as BN-C.
Two-dimensional BC$_2$N nanostructures have also received theoretical attention~\cite{Lu,Lu2}, 
although experimental realization has not been reported yet.

Some types of BN and graphene NRs have been shown, theoretically, to display extraordinary magnetic properties as well,
in that they possess spin polarized edge states: the zigzag graphene NR~\cite{Fujita,Nakada}
is certainly the most well known NR exhibiting this effect,
nevertheless edge magnetism has been predicted to occur in unsaturated BN zigzag NRs~\cite{Barone}
and hexagonal armchair BC$_2$N (BC$_2$N) NRs~\cite{Lu} too.
However, the stability of edge magnetism is still debated~\cite{Wassmann,Kunstmann} and no direct experimental
evidence for the presence of magnetic edge states in these systems has been provided yet.
For deposited NRs, the interaction with the substrate is a serious concern: recently, we have revealed the absence
of edge states in zigzag graphene NRs on Ir$(111)$
by a combined Density Functional Theory (DFT) and scanning tunneling microscopy study~\cite{Dinesh,Yan1}.
We have also investigated zigzag graphene NRs on the $(111)$ surface of Cu, Ag and Au by DFT: our results indicate that
these NRs possess edge states, in agreement with recent experiments~\cite{Tao}, but do not display edge magnetism,
with the exception of H-terminated graphene NRs on Au$(111)$.~\cite{Yan2}

It has been shown experimentally that Cu substrates can be effectively used for the growth of graphene~\cite{Li2}
and BN~\cite{Preobrajenski,Joshi} nanostructures, as well as hybrid BN-C heterostructures~\cite{Ci,Liu}.
Here, we present a DFT study of BN-C, BN and BC$_2$N hexagonal MLs and NRs on the Cu$(111)$ surface, aiming at understanding
the effect of this substrate on the structural and electronic properties of the MLs and NRs.
We fully relax all the models of the MLs and the NRs and determine the most favorable adsorption sites for the
B, N and C atoms.
We show that monolayers with alternating domains interact significantly with Cu$(111)$, due
to the hybridization of the Cu 3d orbitals with states localized at the B-C interfaces.
As a result, the MLs bend considerably so as to form valleys at these interfaces. 
The electronic properties of the MLs are strongly affected by this interaction too.
BN and BC$_2$N MLs, on the other hand, interact weakly with the substrate.
In the case of NRs, we primarily focus on the electronic states and chemical bonding mechanisms responsible for the interaction 
between the NRs and the surface near the edge.
We show that, in the unpassivated case, the latter interaction is determined by a complex interplay between the hybridization of
the d states of the surface Cu atoms with a) the dangling-bond orbitals of the edge atoms and b) the edge states of the NRs. 
For passivated NRs, the interaction is weaker and involves only the edge states of the NRs.
We also show that, although some of the studied models have magnetic edge states and/or interface states in a free-standing configuration, 
none of them exhibits significant magnetization at the edge/interface upon deposition on the Cu$(111)$ surface.

\section{Computational methods}

The structural optimization and the calculation of the electronic properties were carried out
using the plane-wave package Quantum-Espresso~\cite{QE}.
We employed gradient-corrected exchange correlation functionals~\cite{pbe},
semi-empirical van der Waals corrections~\cite{Grimme} and
(scalar-relativistic) ultrasoft pseudopotentials~\cite{Vanderbilt}.
The use of Grimme corrections is crucial, in that plain GGA functionals have been shown to yield
a negative binding energy between BN sheets and Cu$(111)$~\cite{Joshi}.

In the following description of the models, the $z$ axis is taken perpendicular to the surface.
We use the same lattice parameter of 2.5 {\AA} for both BN and graphene. This value corresponds to the 
experimental lattice constant for BN (the experimental parameter for graphene is 2.46 {\AA}). 
To investigate the large supercells required to account for the mismatch between Cu$(111)$ and graphene or BN is computationally
%
%
extremely demanding; for this reason, we used a compressed (about 3.8 \%) Cu lattice to make Cu$(111)$ and the MLs commensurate.
%
%
Therefore, the primitive $(1 \times 1)$ cell of Cu$(111)$ was used for BN MLs, whereas, for BC$_2$N MLs,
a $(1 \times \sqrt{3})$ supercell of Cu$(111)$ was employed.
As regards BN-C heterostructures, we considered two models consisting of a periodic alternation of BN and C stripes.
In the first model (which we denote BN-C$^{(1)}$ in the following), 
BN and C stripes have a width of 4 units of the hexagonal lattice, 
whereas in the second model (denoted BN-C$^{(2)}$),
the width is equal to 8 units (see also Figure~\ref{struct_MLs}).
Although the widths of the C and BN domains in our models are much smaller than those reported 
in recent experiments~\cite{Liu}, the models can nevertheless provide important information about the interaction
of the BN-C heterostructures with the Cu substrate near the interface between different domains.
The $x$ axis was taken parallel to the stripes and the $y$ axis in the plane and normal to them.
A $(1 \times 8\sqrt{3}))$ Cu$(111)$ supercell was used for both models.
The Cu$(111)$ surface was modeled with slabs consisting of four atomic layers. 
To separate the periodic images of the models along the $z$ direction, vacuum layers with thicknesses of 9 {\AA} at least
were used.
In Ref.~\onlinecite{Yan2} some test calculations employing thicker slabs 
(containing up to 12 layers) were performed for graphene NRs on Cu$(111)$.
These calculations showed that 4-layer slabs are sufficient to describe
the interaction between the NRs and the Cu surface, which makes us confident that the same holds true for the
models studied in this work.
We considered a zigzag termination for all the models of the NRs except the BC$_2$N systems, for which the armchair termination
was studied. Both H-free and singly H-terminated NRs were investigated.
We considered BN and BC$_2$N NRs with a width of 8 unit cells of the honeycomb lattice.
The $x$ axis was taken parallel to the NRs.
$(1 \times 5\sqrt{3})$ and $(\sqrt{3} \times 5\sqrt{3})$ supercells of Cu$(111)$  
were employed for BN and BC$_2$N NRs respectively, corresponding to distances between 
nearest-neighbor periodic images of the NRs of at least 14 {\AA}.  
Two models of BN-C NRs were also considered. Both models were ended with BN stripes at the two edges. The two NRs consist 
of three stripes (BN-C-BN) and five stripes (BN-C-BN-C-BN) respectively. The widths of the
stripes at the edges are equal to four and two units of the honeycomb lattice, whereas the widths of the stripes in the
center coincide with those of the two models of BN-C MLs discussed above.
$(1 \times 8\sqrt{3}))$ Cu$(111)$ supercells were used for these models, 
corresponding to a distance between periodic images of at least 16 {\AA}.
$14 \times 1 \times 1$ Monkhorst-Pack meshes~\cite{MP} were used to perform the integration over the Brillouin zone for all
the models of NRs.\\
All of the atoms of the NRs and the MLs, as well as the two topmost Cu layers,
were allowed to relax during structural optimization.\\
In the following sections, we assume that the Fermi energy, E$_{\rm F}$, is at zero energy. 

\section{Results}

\subsection{Monolayers of BN, BN-C and BC$_2$N on Cu$(111)$}

\begin{figure}[t!]
\begin{center}
\includegraphics[width=0.8\columnwidth]{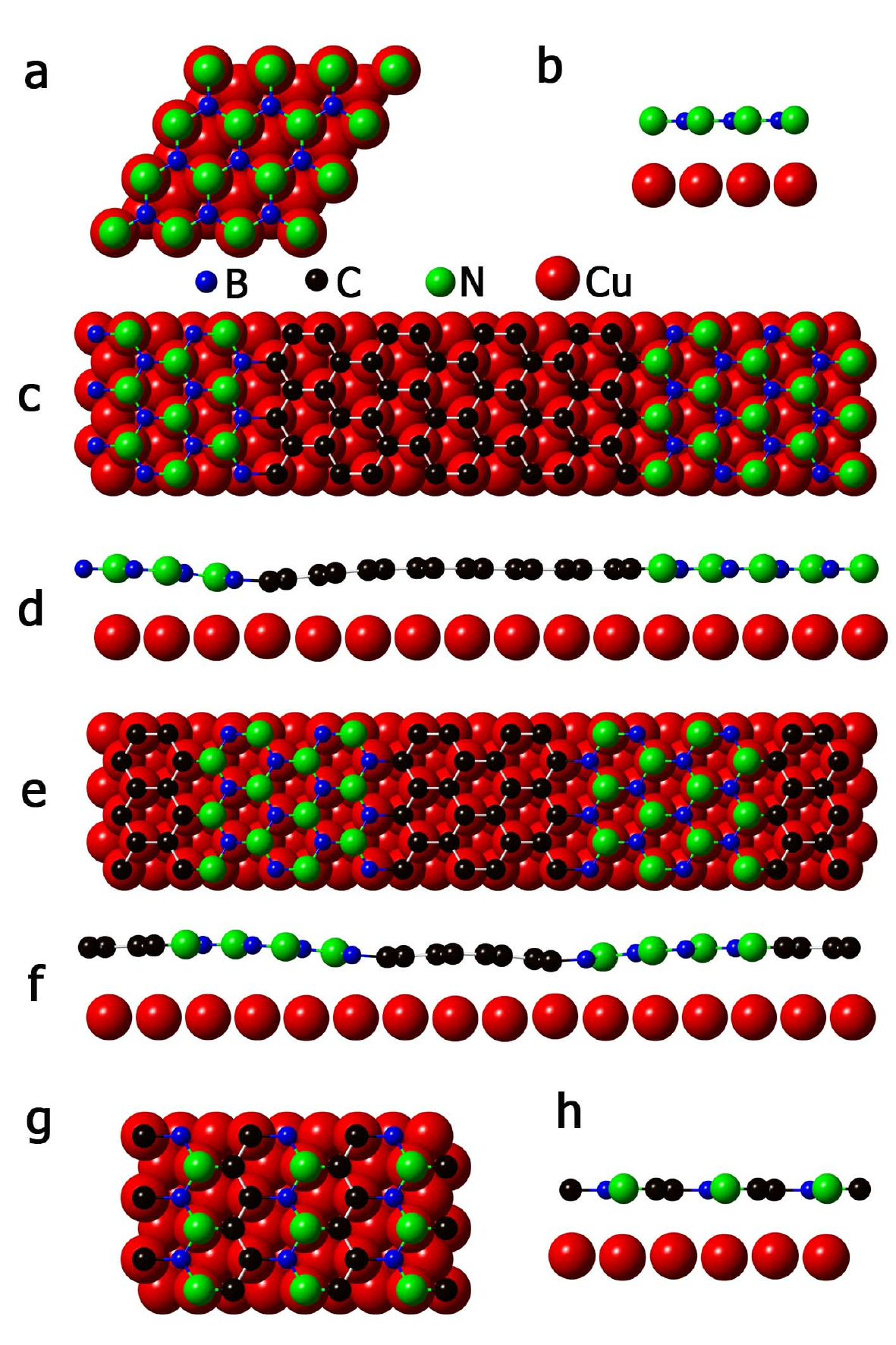}
\end{center}
\caption{
Topography of the relaxed models of BN, BN-C and BC$_2$N monolayers on Cu$(111)$.
For the sake of clarity, here and in the following figures only the top Cu layer is shown.
(a-b) Top and side view of the BN ML on Cu$(111)$.
(c-d) Top and side view of the BN-C$^{(1)}$ ML on Cu$(111)$.
(e-f) Top and side view of the BN-C$^{(2)}$ ML on Cu$(111)$.
(g-h) Top and side view of the BC$_2$N ML on Cu$(111)$.}
\label{struct_MLs}
\end{figure}

Top and side views of the models of BN, BN-C and BC$_2$N monolayers on the Cu$(111)$ surface are shown in Fig.~\ref{struct_MLs}.
We considered several configurations with different adsorption sites for the B, C and N atoms and optimized their geometry.
Here we discuss the most energetically favorable configurations only.
In the BN case, the most stable sites for N and B atoms are ontop and hollow respectively, as already discussed in
Ref.~\onlinecite{Joshi}.
In the case of BC$_2$N sheets, B and N atoms sit at ontop and hollow sites too, whereas C atoms occupy both ontop
and hollow sites.
The interaction between these MLs and the Cu$(111)$ surface is weak: the distance between BN (resp. BC$_2$N) sheets
and the surface is equal to 2.94 (2.86) {\AA}.
Here, it is important to stress that the use of semi-empirical van der Waals corrections (DFT-D2) can only partially cure
the inability of standard GGA functionals to describe dispersive interactions accurately.
In fact, these corrections seem to lead to slight overbinding in the case of BN MLs on Cu$(111)$~\cite{Joshi}.
Hence, the obtained values of the ML-Cu$(111)$ distances might deviate from the experimental ones.
The band gaps of the BN and BC$_2$N MLs are equal to 4.63 eV and 1.62 eV respectively.

\begin{figure}[t!]
\begin{center}
\includegraphics[width=0.6\columnwidth]{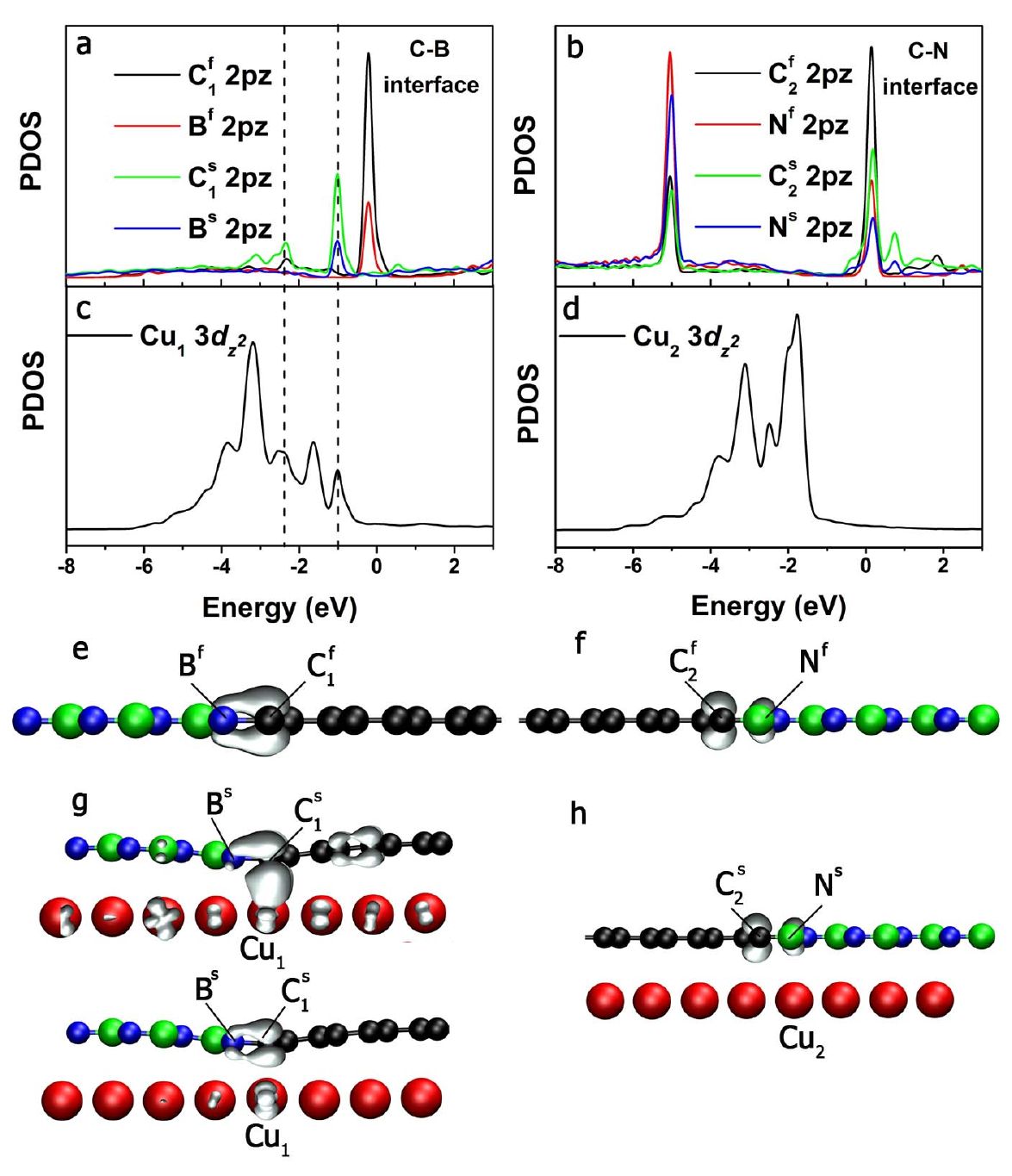}
\end{center}
\caption{
Electronic properties of the free-standing and deposited BN-C$^{(1)}$ monolayer.
Since the BN-C$^{(2)}$ monolayer has very similar electronic properties, it is not shown here.
(a) Non-spin-polarized 2p$_z$-PDOS of the B and C atoms at the B-C interface of the free-standing (B$^f$, C$_1^f$) and
    supported ML (B$^s$, C$_1^s$).
(b) Non-spin-polarized 2p$_z$-PDOS of the N and C atoms at the C-N interface of the free-standing (N$^f$, C$_2^f$) and
    supported ML (N$^s$, C$_2^s$).
(c-d) 3d$_{z^2}$-PDOS of the Cu atom beneath the B-C interface (Cu$_1$, Fig.~c)
    and the C-N interface (Cu$_2$, Fig.~d).
(e) Plot of a charge isosurface of the localized state at the B-C interface of the free-standing ML. The state
    corresponds to the PDOS peaks at E$_{\rm F}$ in Fig.~a. The state has bonding character.
(f) Plot of a charge isosurface of the localized state at the C-N interface of the free-standing ML. The state
    corresponds to the PDOS peaks at E$_{\rm F}$ in Fig.~b. The state has antibonding character.
(g) Plot of charge isosurfaces of the two (bonding and antibonding) states resulting from the hybridization of the B-C
    interface state with the Cu d$_{z^2}$ orbital. The states correspond to the PDOS peaks in Fig.~a
    at -2.3 eV and -1.0 eV respectively.
(h) Plot of a charge isosurface of the localized state at the C-N interface of the deposited ML. The state
    barely interacts with the substrate.}
\label{PDOS_MLs}
\end{figure}

The structure of deposited BN-C MLs is more complicated, which stems from the fact that the free-standing MLs possess localized 
(spin-polarized) states at the interface between BN and graphene domains. 
In Figs.~\ref{PDOS_MLs}(a)-(b) the non-spin-polarized projected DOS (PDOS) onto 2p$_z$ orbitals of B, C and N atoms are shown: the peaks
at E$_{\rm F}$ correspond to the interface states of the free-standing MLs. These states hybridize with the Cu d states.
More precisely, the interaction is strong at B-C interface (see the side view of the two models in Fig.~\ref{struct_MLs}).
This behaviour can be understood by considering the different properties of the states localized at the B-C and C-N interfaces.
In the case of the B-C interface, the localized state is a bonding state
between the p$_z$ orbitals of the B and C atoms (Fig.~\ref{PDOS_MLs}(e)), whereas the C-N interface state is an antibonding state
between the N and C p$_z$ orbitals (Fig.~\ref{PDOS_MLs}(f))~\cite{Nakamura}.
The B-C interface state hybridizes with the d orbitals of the Cu atoms beneath, forming bonding and antibonding states
(Figs.~\ref{PDOS_MLs}(c,g)), hence the strong interaction.
On the other hand, the C-N interface state barely interacts with the substrate (Figs.~\ref{PDOS_MLs}(d,h)).
%
%
%
As a result, the two BN-C sheets are not planar: the minimum distance between the MLs and the surface at the B-C interface
is equal to 2.36 {\AA} for both models,
whereas the maximum distance occurs at the center of the graphene (model BN-C$^{(1)}$) and BN (model BN-C$^{(2)}$) stripes
and is equal to 2.99 and 3.13 {\AA} respectively.

The calculated work functions of the MLs deposited on Cu$(111)$ (and of clean Cu$(111)$) and the values of the electron charge
displacement between the MLs and the surface are provided in Table~\ref{WFs}.
Charge transfers were calculated based on the Bader charge analysis~\cite{Tang}.

\begin{table}
\begin{tabular}{ccc}
\hline \hline                       &       &                      \\
\hline
$\;\;\;\;\;\;\;\;\;\;\;\;\;\;\;\;$  &  WF (eV)  &         CD (e/atom)          \\
\hline
Cu$(111)$                           & $\;\;$  4.84 $\;\;$ & $\;\;$         -            $\;\;$ \\
BN on Cu$(111)$                     & $\;\;$  3.67 $\;\;$ & $\;\;$ $3 \cdot 10^{-3} $  $\;\;$ \\
BC$_2$N on Cu$(111)$                   & $\;\;$  3.70 $\;\;$ & $\;\;$ $8 \cdot 10^{-3} $ $\;\;$ \\
BN-C$^{(1)}$ on Cu$(111)$           & $\;\;$  3.78 $\;\;$ & $\;\;$ $12 \cdot 10^{-3} $ $\;\;$ \\
BN-C$^{(2)}$ on Cu$(111)$           & $\;\;$  3.74 $\;\;$ & $\;\;$ $13 \cdot 10^{-3} $ $\;\;$ \\
\hline \hline
\end{tabular}
\caption{Work function (WF) of the bare Cu$(111)$ and of the Cu$(111)$ with deposited MLs and charge displacement (CD) per ML atom 
between the ML and the surface. Positive values of CD indicate electron charge displacement to the ML.}
\label{WFs}
\end{table}
The work functions of Cu$(111)$ and of the BN ML on Cu$(111)$ compare well with previous theoretical work
(Ref.~\onlinecite{DaSilva} and \onlinecite{Joshi} respectively). Generally, the presence of the MLs leads to a decrease
of the Cu$(111)$ work function of the order of 1.1-1.2 eV.
As far as the electronic charge distribution at the interface is concerned, 
it turns out that a charge displacement to the ML occurs for all the deposited models.
In the case of BN and BC$_2$N, the charge displacement at the interface does not lead to 
n-doping of these two insulating MLs. 
The rearrangement is due to polarization effects originating from the presence of the substrate.  
A more significant charge transfer occurs in the case of the BN-C models: as a result, these MLs are n-doped.
Perfect monolayer graphene has also been shown to be n-doped when deposited on Cu$(111)$~\cite{Giovannetti}:
the corresponding charge transfer, $11 \cdot 10^{-3} $ electrons per atom 
($8 \cdot 10^{-3} $ electrons per atom in the case of LDA calculations~\cite{Khomyakov}), 
is slightly smaller than for BN-C MLs.
The phenomenological model introduced in Ref.~\onlinecite{Giovannetti} to explain doping and work function trends for graphene
physisorbed on metallic substrates (including Cu, Ag, Pt and Au) cannot be extended to the case of BN-C on Cu$(111)$, due to 
the stronger chemical interaction between the two systems. 

The two BN-C MLs are magnetic in the free-standing case, owing to the spin-polarized states localized at the B-C
and C-N interfaces. However, they become non-magnetic upon deposition onto the Cu substrate
(the magnetization is less than $0.01 \mu_B$ per interface atom). The demise of magnetism originates from
a) the said chemical bonding at the B-C interface, as a result of which bonding and antibondings states
are formed between the interface states and Cu d states, which are both occupied, and b) the emptying of the interface states
at the C-N interface (the corresponding PDOS peaks are shifted by 0.2 eV).

\subsection{BN nanoribbons on Cu$(111)$}

We considered both H-free and H-terminated BN NRs on the Cu$(111)$ surface.
Zigzag BN NRs with unpassivated edges have been shown to possess spin-polarized edge states
due to the dangling bonds of the edge atoms~\cite{Barone}.
However, this configuration is expected to be very reactive and magnetism disappears if the edges
are passivated with H atoms.
In fact, the singly H-terminated BN NR has an occupied, non-spin-polarized edge state localized at the N edge
and an unoccupied state localized at the B edge~\cite{Nakamura}. The energy gap of this system corresponds to
the energy difference between these two states (albeit at different k-points~\cite{Nakamura}). 
Interestingly, this gap becomes smaller than the band gap of perfect
monolayer BN for sufficiently wide NRs due to a self-doping effect, namely an enhanced charge transfer from
the B edge to the N edge~\cite{Nakamura}.

\begin{figure}[t!]
\begin{center}
\includegraphics[width=0.8\columnwidth]{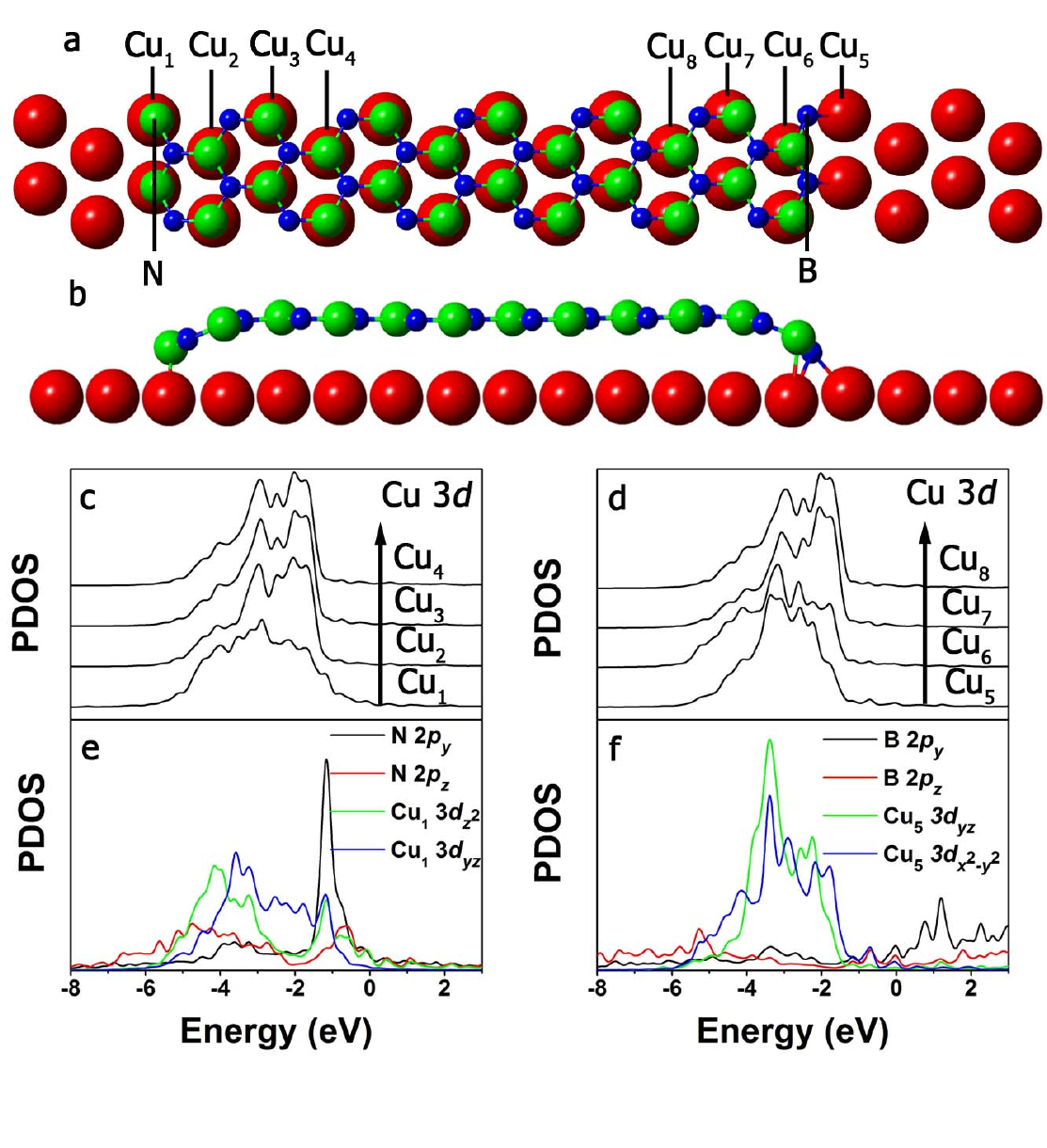}
\end{center}
\caption{
Structural and electronic properties of a H-free zigzag BN NR on Cu$(111)$.
(a-b) Topography of the relaxed model.
    Cu atoms are labeled by numbers indicating different chemical environments and used in (c)-(f).
(c-d) PDOS of the 3d states of several Cu atoms starting from an atom
    below the left (c) or right (d) edge of the NR towards an atom below the centre of the NR.
(e) PDOS of the 2p$_y$ and 2p$_z$ orbitals of a N atom at the left edge of the NR, sitting at on-top site, and
    PDOS of the 3d$_{yz}$ and 3d$_{z^2}$ states of the nearest neighbor Cu atom (Cu$_{1}$).
(f) PDOS of the 2p$_y$ and 2p$_z$ orbitals of a B atom at the right edge of the NR, sitting at hollow site, and
    PDOS of the 3d$_{yz}$ and 3d$_{x^2-y^2}$ states of a nearest neighbor Cu atom (Cu$_{5}$).}
\label{PDOS_noH_BN}
\end{figure}

We first discuss H-free NRs on Cu$(111)$. We considered several configurations
with different adsorption sites for the edge B and N atoms.
The lowest energy structure is shown in Figs.~\ref{PDOS_noH_BN}(a)-(b): in this configuration,
the edge B and N atoms sit at hollow and on-top sites respectively. In fact, all the B (resp. N) atoms of the NR
sit at hollow or quasi-hollow (resp. on-top or quasi on-top) sites, analogously to the case of BN MLs.
The strong interaction between the edge atoms and the Cu surface leads to the bending of the NR at both sides:
as shown in Table I, the distance between the B (N)  atoms and the nearest-neighbor Cu atoms
is 1.92 {\AA} (2.09 {\AA} ) at the edge, whereas the maximum distance between the NR and Cu$(111)$
is 3.08 {\AA} (see also Fig.~\ref{PDOS_noH_BN}(b)).
\begin{table}
\begin{tabular}{ccc}
\hline \hline                       &        &               \\
\hline
$\;\;\;\;\;\;\;\;\;\;\;\;\;\;\;\;$  &   min  &   max  \\
\hline
BN              &  $\;\;$ 1.92 (N); 2.09 (B)             $\;\;$  & $\;\;$ 3.08 $\;\;$ \\
H-BN            &  $\;\;$ 2.82 (N); 2.28 (B)             $\;\;$  & $\;\;$ 3.04 $\;\;$ \\
BN-C$^{(1)}$    &  $\;\;$ 1.95 (N); 2.09 (B); 2.37 (B-C) $\;\;$  & $\;\;$ 3.11 $\;\;$ \\
H-BN-C$^{(1)}$  &  $\;\;$ 2.84 (N); 3.07 (B); 2.37 (B-C) $\;\;$  & $\;\;$ 3.05 $\;\;$ \\
BN-C$^{(2)}$    &  $\;\;$ 2.00 (N); 2.01 (B); 2.44 (B-C) $\;\;$  & $\;\;$ 3.38 $\;\;$ \\
H-BN-C$^{(2)}$  &  $\;\;$ 3.20 (N); 2.94 (B); 2.49 (B-C) $\;\;$  & $\;\;$ 3.07 $\;\;$ \\
BC$_2$N            &  $\;\;$ 2.08 (B); 2.09 (C)             $\;\;$  & $\;\;$ 3.29 $\;\;$ \\
H-BC$_2$N          &  $\;\;$ 2.47 (B); 2.17 (C)             $\;\;$  & $\;\;$ 3.04 $\;\;$ \\
\hline \hline
\end{tabular}
\caption{Minimum distance between the edge atoms of the NRs and the nearest neighbor atoms of
the Cu$(111)$ surface and maximum distance between the NR and the surface.
(1) and (2) stand for the BN-C-BN and BN-C-BN-C-BN models respectively.
In the case of BN-C NRs, the minimum distance between
Cu atoms and the NRs at the B-C interface is provided as well. Distances are in Angstrom.}
\label{distances}
\end{table}
Since the NR is not parallel to the surface at the edge, the dangling-bond orbitals of the B and N atoms
are not pure 2p$_y$ orbitals but linear combinations of 2p$_z$ and 2p$_y$ orbitals, albeit with predominant p$_y$ character.
Analogously, the edge states have predominant, but not exclusive, p$_z$ character.
Figs.~\ref{PDOS_noH_BN}(c)-(d) display the PDOS of d states of several Cu atoms,
starting from an atom below the edge of the NR towards an atom
below the centre.
The strong hybridization at the edge leads to the formation of new peaks in the Cu PDOS.
The PDOS onto 2p orbitals of edge N and B atoms and onto
3d orbitals of the neighboring Cu atoms are shown in Figs.~\ref{PDOS_noH_BN}(e)-(f).
The chemical interaction at the N edge bears some similarities with that at the edges of H-free graphene NRs deposited
on Cu, Ag and Au $(111)$ surfaces~\cite{Yan2}.
As evidenced from Fig.~\ref{PDOS_noH_BN}(e), the dangling orbital of the edge N atoms forms bonding and antibonding states with the d orbitals
(in particular, d$_{yz}$ and d$_{z^2}$) of the nearest neighbor Cu atom. These states correspond to the broad peak centered
around -3.7 eV and the sharp peak at -1.23 eV respectively.
The main peak of the N edge state is centered at -0.7 eV. This state hybridizes mainly with the d$_{z^2}$ orbital of the Cu atom.

At the B edge, the dangling-bond orbital of the B atoms hybridizes with the
d$_{yz}$ and d$_{x^2-y^2}$ orbitals of the neighboring Cu atoms (Fig.~\ref{PDOS_noH_BN}(d)). The B edge state is unoccupied and does not
play a role in the bonding.

\begin{figure}[t!]
\begin{center}
\includegraphics[width=1.0\columnwidth]{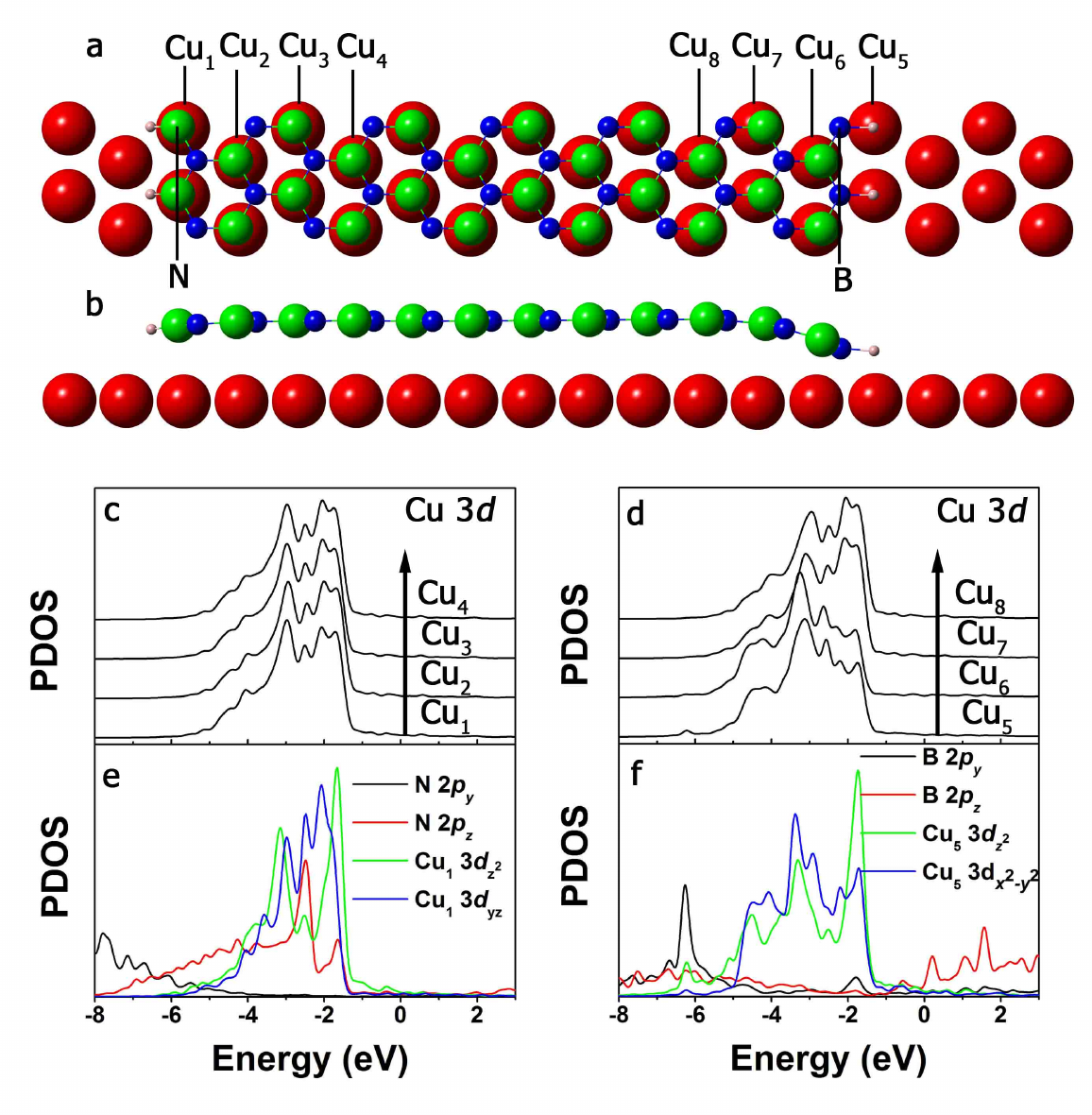}
\end{center}
\caption{
Structural and electronic properties of a H-terminated zigzag BN NR on Cu$(111)$.
(a-b) Topography of the relaxed model.
    Cu atoms are labeled by numbers indicating different chemical environments and used in (c)-(f).
(c-d) PDOS of 3d states of several Cu atoms starting from an atom
    below the left (c) or right (d) edge of the NR towards an atom below the centre of the NR.
(e) PDOS of the 2p$_y$ and 2p$_z$ orbitals of a N atom at the left edge of the NR, sitting at on-top site, and
    PDOS of the 3d$_{yz}$ and 3d$_{z^2}$ states of the nearest neighbor Cu atom (Cu$_{1}$).
(f) PDOS of the 2p$_y$ and 2p$_z$ orbitals of a B atom at the right edge of the NR, sitting at hollow site, and
    PDOS of the 3d$_{yz}$ and 3d$_{z^2}$ states of a nearest neighbor Cu atom (Cu$_{5}$).}
\label{PDOS_H_BN}
\end{figure}

Next, we consider H-terminated BN NRs on Cu$(111)$. In the lowest energy configuration shown in Figs.~\ref{PDOS_H_BN}(a)-(b),
the edge N and B atoms sit at on-top and quasi hollow site, similarly to the H-free NR and the ML case.
At the N edge, no significant bending occurs (Fig.~\ref{PDOS_H_BN}(b)).
The N edge state hybridizes with the d$_{z^2}$ orbital of the Cu atoms beneath,
forming bonding and antibonding states located around $-2.42$ and $-1.59$ eV (see Fig.~\ref{PDOS_H_BN}(e)).\\
Somewhat surprisingly, the interaction with the surface is stronger at the B edge (see Table 2 and Fig.~\ref{PDOS_H_BN}(b)).
The edge B atom forms an additional bond with one of the Cu atoms beneath
(denoted Cu$_5$ in Fig.~\ref{PDOS_H_BN}(f)):
the distance between these two atoms is equal to 2.28 {\AA}.
To understand the bonding mechanism in this model, it is useful to consider a toy model consisting of a
free-standing BN ribbon with a doubly H-terminated B edge. The presence of a second H atom in this model
leads to charge transfer from the nearest neighbor N atom to B.
As a result, an edge state localized at the N, B and H atoms appears
at the Fermi energy of the ribbon. Qualitatively, a similar effect should occur for the deposited H-terminated NR,
due to the additional bond between edge B and Cu$_5$, however
here the edge state interacts with the Cu surface
(the distance between the N atom at the second row and the Cu atom beneath is only 2.27 {\AA}),
resulting in a bonding state with an energy of about -1.7 eV (Fig.~\ref{PDOS_H_BN}(f)).
Hence, this configuration leads to a strong chemical interaction between the NR and Cu$(111)$.
We found another configuration where N and B atoms also sit at on-top and hollow sites but the interaction
at the B edge is weaker and, therefore, B-Cu distances are larger (see supplement). In this configuration, 
the distance between the edge B atom and the nearest-neighbor Cu atom is 2.99 {\AA}. 
This model has a slightly higher energy (0.05 eV per edge B atom) than the one discussed above.\\
Both configurations do not exhibit magnetization at the edge (less than $10^{-3} \mu_B$ per edge atom).

\subsection{BN-C nanoribbons on Cu$(111)$}

\begin{figure}[t!]
\begin{center}
\includegraphics[width=1.0\columnwidth]{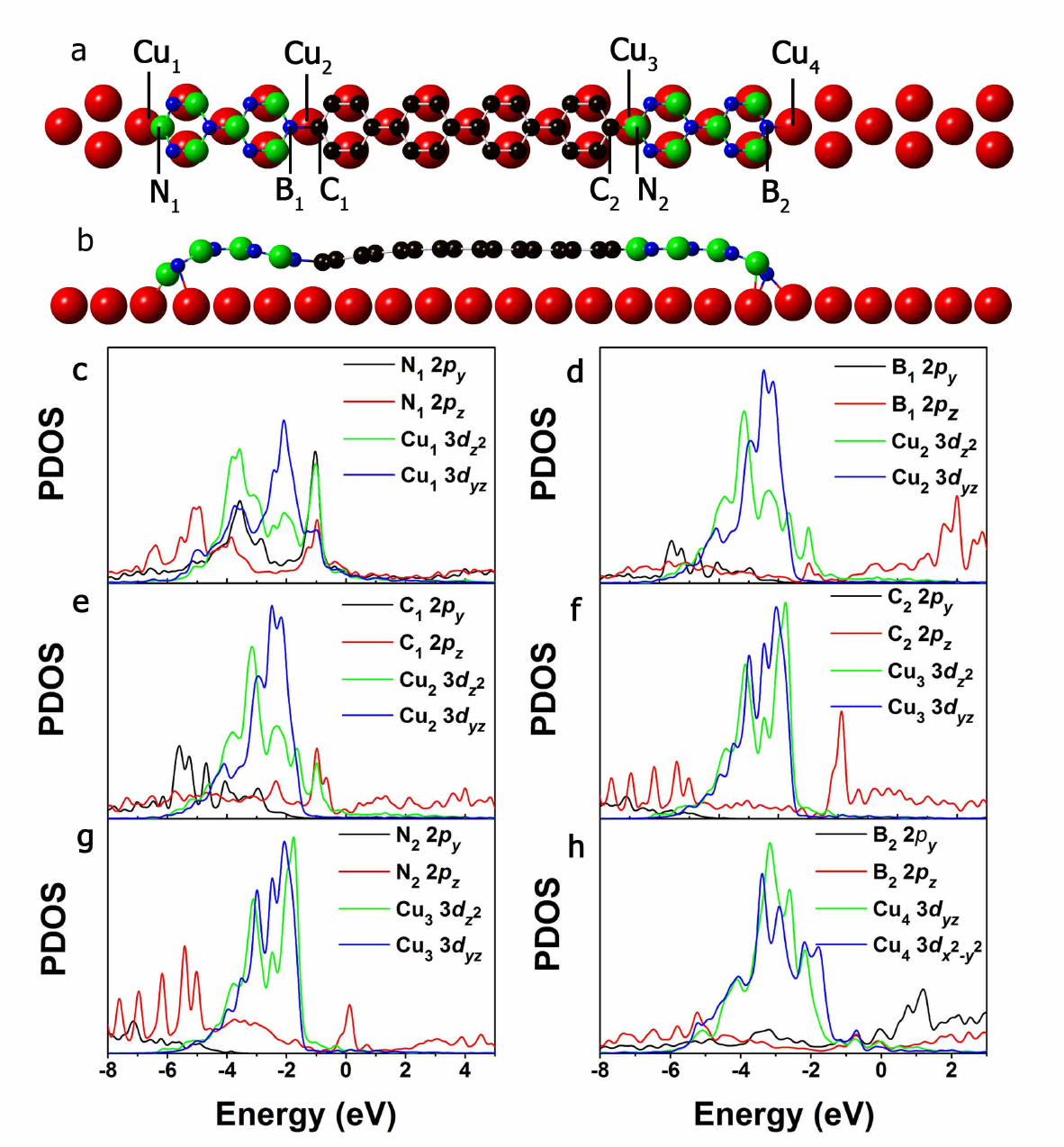}
\end{center}
\caption{
Structural and electronic properties of a H-free zigzag BN-C-BN NR on Cu$(111)$.
(a-b) Topography of the relaxed model.
    B, C, N and Cu atoms are labeled by numbers indicating different chemical environments and used in (c)-(h).
(c-h) PDOS of the 2p$_y$ and 2p$_z$ orbitals of B, C and N atoms at the edges or interfaces of the NR and
    PDOS of some 3d states of the corresponding nearest neighbor Cu atoms.}
\label{PDOS_noH_BN-C-BN}
\end{figure}

\begin{figure}[t!]
\begin{center}
\includegraphics[width=1.0\columnwidth]{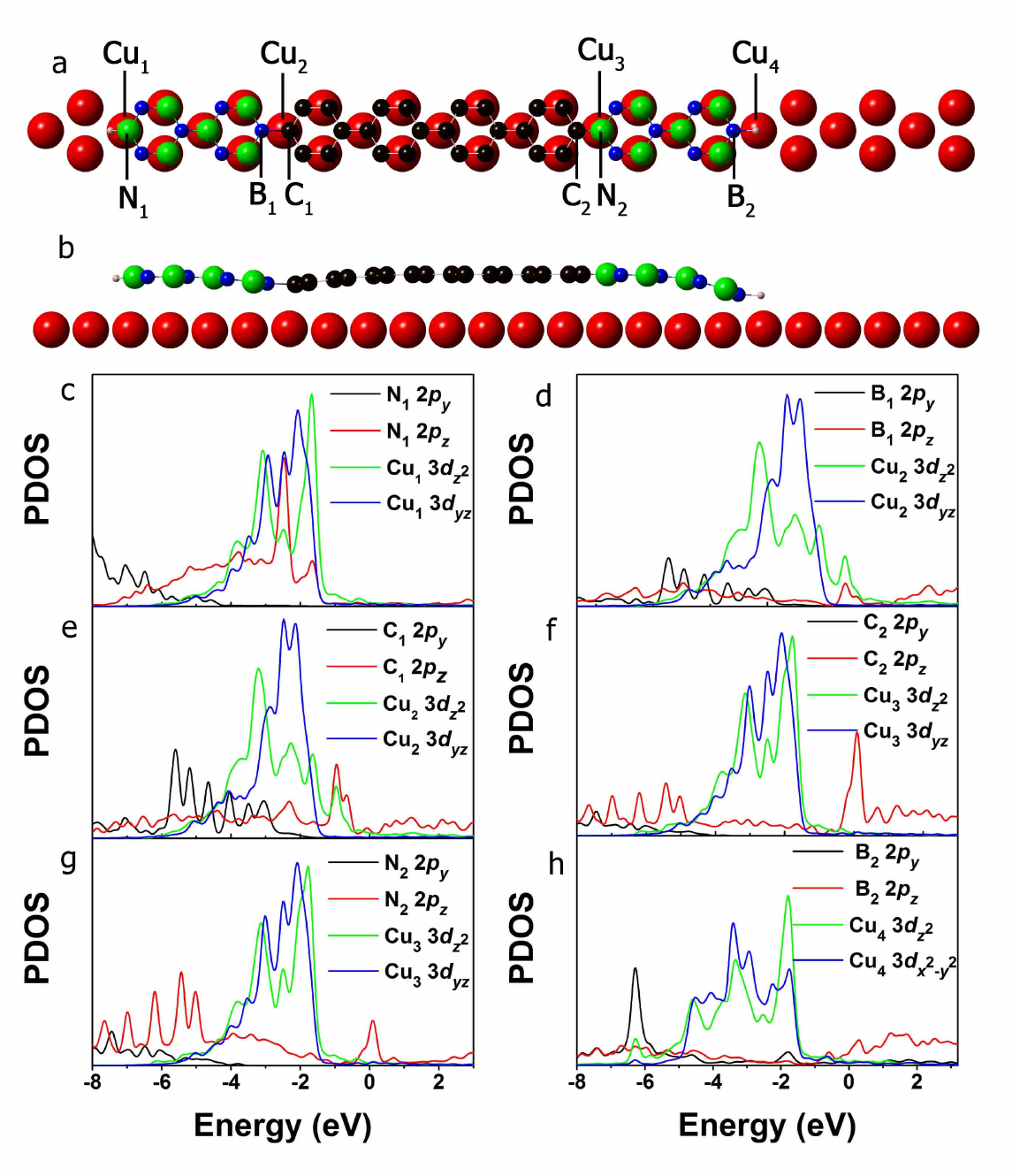}
\end{center}
\caption{
Structural and electronic properties of a H-terminated zigzag BN-C-BN NR on Cu$(111)$.
(a-b) Topography of the relaxed model. 
    B, C, N and Cu atoms are labeled by numbers indicating different chemical environments and used in (c)-(h).
(c-h) PDOS of the 2p$_y$ and 2p$_z$ orbitals of B, C and N atoms at the edges or interfaces of the NR and
    PDOS of some 3d states of the corresponding nearest neighbor Cu atoms.}
\label{PDOS_H_BN-C-BN}
\end{figure}

We consider first the BN-C NRs made of three alternating BN and C stripes shown
in Figs.~\ref{PDOS_noH_BN-C-BN}-\ref{PDOS_H_BN-C-BN}.
The central graphene stripe is made of eight units of the hexagonal lattice, while
the two edge BN stripes consist of four units.
Obviously, there are several similarities between the bonding mechanism at the edges in these two models (H-free and
H-terminated) and in the corresponding models of the BN NRs.
More specifically, the B and N atoms at the edge sit at hollow (or quasi hollow) and on-top (or quasi on-top) sites.
Furthermore, in the H-free case, the dangling orbital of the edge N atoms forms bonding and antibonding states
with the d$_{yz}$ and d$_{z^2}$ orbitals of the Cu atom beneath (Fig.~\ref{PDOS_noH_BN-C-BN}c).
The edge state hybridizes mainly with the d$_{z^2}$ orbital of the same Cu atom.
At the B edge, the dangling-bond orbital of the B atoms hybridizes with the
d$_{yz}$ and d$_{x^2-y^2}$ orbitals of the three nearest neighbor Cu atoms (Fig.~\ref{PDOS_noH_BN-C-BN}h).\\
In the H-terminated case, the B edge interacts more strongly with the surface than the N edge
(Figs.~\ref{PDOS_H_BN-C-BN}(a-c,h)). 
The discussion presented in the previous section
about the bonding mechanisms at both edges is valid for this model as well.
The properties of the interface states localized at the BN-C interfaces and their interaction
with the Cu substrate (Figs.~\ref{PDOS_noH_BN-C-BN}(d-g) and \ref{PDOS_H_BN-C-BN}(d-g))
are completely analogous to the case of the corresponding BN-C ML on Cu$(111)$.

Similar considerations hold for the BN-C NRs made of five alternating BN and C stripes, 
as discussed in the supplement~\cite{supplement}.
All of the NRs considered in this section are magnetic in the free-standing case, due to the 
spin-polarized interface states already discussed in the ML case. Similarly to the latter systems,
interface magnetism completely disappears in the presence of the Cu substrate.

\subsection{BC$_2$N nanoribbons on Cu$(111)$}

It was recently shown that free-standing, H-terminated armchair BC$_2$N NRs possess magnetic edge states
when the edges are ended with C and B atoms (or C and N atoms)~\cite{Lu}.
In both cases, the p$_z$ orbitals of the edge C atoms mainly contribute to the charge density of the edge state.
The contribution of the B (or N) p$_z$ orbitals is smaller, but still significant.
In these systems, edge magnetism is stabilized by a self-doping effect and the magnitude of the spin polarization
increases by increasing the NR width.
The order is ferromagnetic along the NR in both cases.
Interestingly, the magnetic order across the NRs depends on the termination:
it is ferromagnetic in the case of B-C termination but antiferromagnetic 
if the edges are ended with C and N atoms~\cite{Lu}.

\begin{figure}[t!]
\begin{center}
\includegraphics[width=1.0\columnwidth]{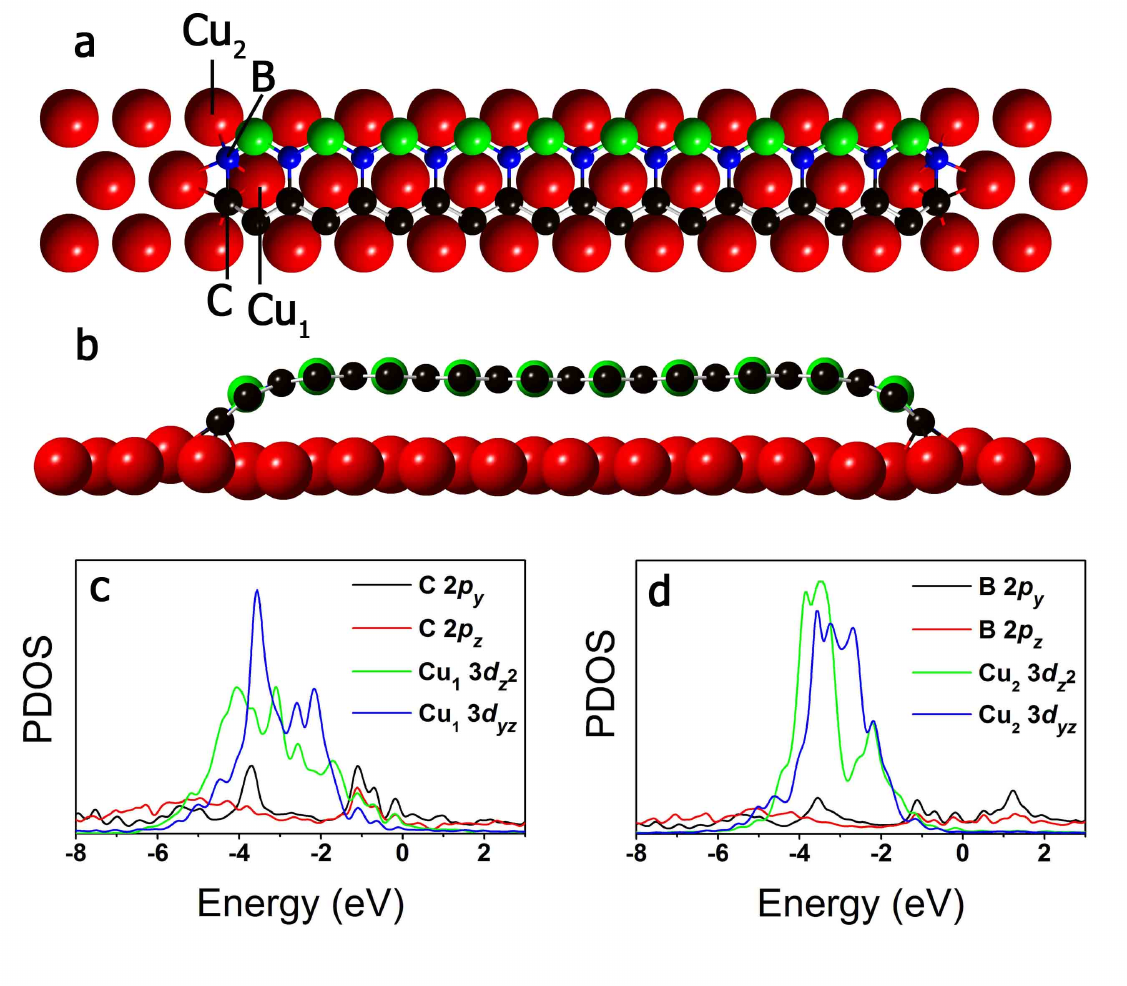}
\end{center}
\caption{
Structural and electronic properties of a H-free armchair BC$_2$N NR on Cu$(111)$.
(a-b) Topography of the relaxed model.
    Cu atoms are labeled by numbers indicating different chemical environments and used in (c)-(d).
(c) PDOS of the 2p$_y$ and 2p$_z$ orbitals of a C atom at the edge of the NR and
    PDOS of the 3d$_{yz}$ and 3d$_{z^2}$ states of the nearest neighbor Cu atom (Cu$_{1}$).
(d) PDOS of the 2p$_y$ and 2p$_z$ orbitals of a B atom at the edge of the NR and
    PDOS of the 3d$_{yz}$ and 3d$_{z^2}$ states of the nearest neighbor Cu atom (Cu$_{2}$).}
\label{PDOS_noH_BC2N}
\end{figure}

In the following, we consider only BC$_2$N NRs with B-C edges. The lowest-energy configuration for H-free NRs on
Cu$(111)$ is shown in Figs.~\ref{PDOS_noH_BC2N}(a-b). The structure is perfectly symmetric:
the B and C atoms at the two edges sit at quasi bridge sites.
The bending of this NR is more pronounced than that of the H-free BN NR and the maximum distance between the NR
and Cu$(111)$ is 0.21 {\AA} larger.
The dangling-bond orbitals of the B and C atoms and the edge state hybridize with all of the d states of the
nearest neighbor Cu atoms, in particular with the d$_{z^2}$ and d$_{yz}$ orbitals (shown in Figs.~\ref{PDOS_noH_BC2N}(a,c-d)).

\begin{figure}[t!]
\begin{center}
\includegraphics[width=1.0\columnwidth]{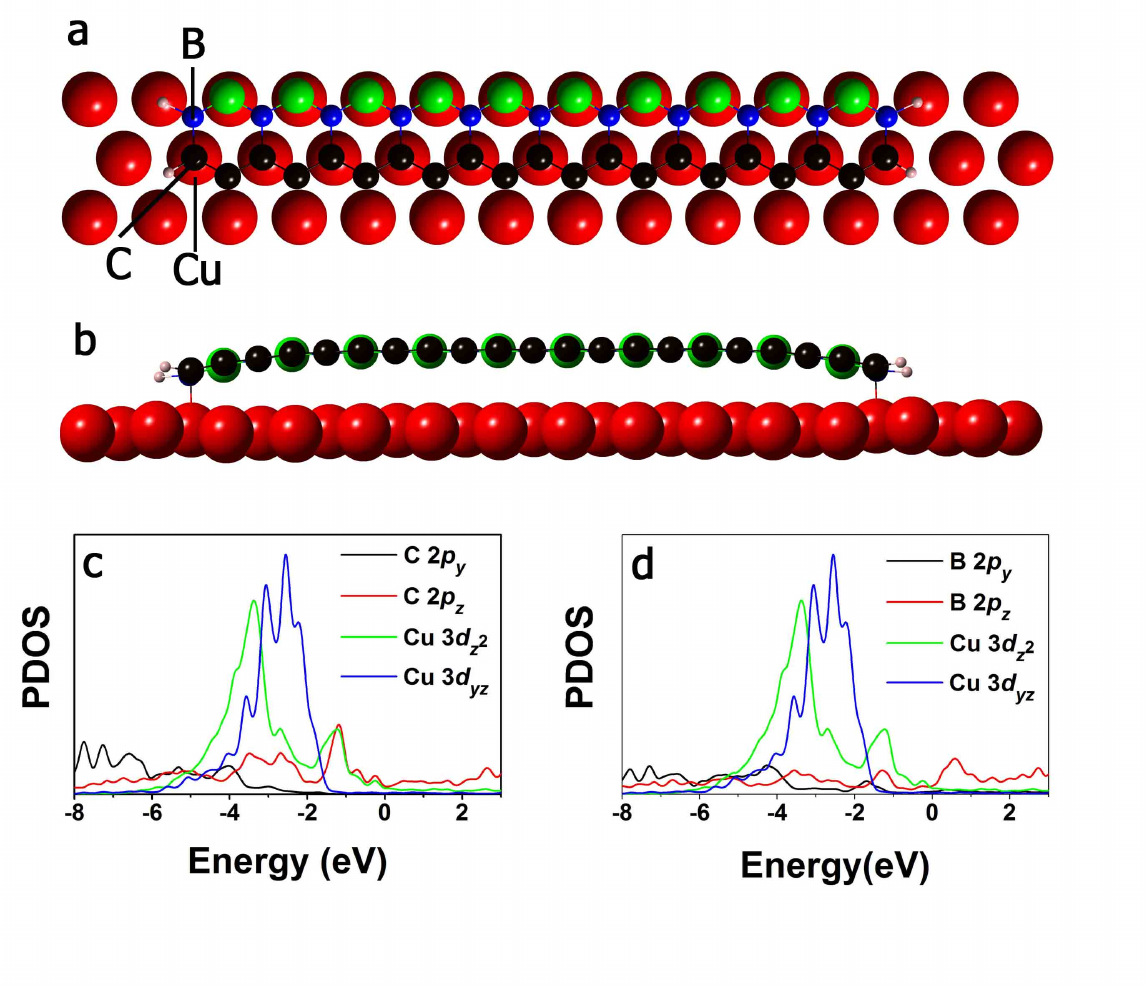}
\end{center}
\caption{
Structural and electronic properties of a H-terminated armchair BC$_2$N NR on Cu$(111)$.
(a-b) Topography of the relaxed model.
(c) PDOS of the 2p$_y$ and 2p$_z$ orbitals of a C atom at the edge of the NR and
    PDOS of the 3d$_{yz}$ and 3d$_{z^2}$ states of the nearest neighbor Cu atom (Cu$_{1}$).
(d) PDOS of the 2p$_y$ and 2p$_z$ orbitals of a B atom at the edge of the NR and
    PDOS of the 3d$_{yz}$ and 3d$_{z^2}$ states of the nearest neighbor Cu atom (Cu$_{2}$).}
\label{PDOS_H_BC2N}
\end{figure}

The most energetically favorable model of H-passivated BC$_2$N NR on Cu$(111)$ is shown in Figs.~\ref{PDOS_H_BC2N}(a-b).
This model is also symmetric:
at both edges, B and C atoms sit at hollow and on-top sites respectively.
The plots of the PDOS of the atoms contributing to the chemical bonding
at the edge are shown in Figs.~\ref{PDOS_H_BC2N}(c-d).
These plots clearly indicate that the edge state of the NR hybridizes mostly with the
d$_{z^2}$ orbital of the Cu atom beneath the C atom.
Both models become non magnetic when deposited on Cu$(111)$.

\section{Conclusions}

We have investigated the structural and electronic properties of BN, BC$_2$N and hybrid BN-C MLs and NRs deposited
on the Cu$(111)$ surface by first-principles simulations. We have shown that all of the MLs decrease the work function of the Cu surface
in a similar way. The interaction of the BN and BC$_2$N MLs with the surface is weak, whereas the BC-N models
consisting of BN and graphene stripes interact significantly with the surface at the B-C interfaces, where
localized interface states exist.
This interaction leads to a deformation of the MLs and a relatively large charge transfer between the BC-N models 
and Cu$(111)$, resulting in significant n-doping of the MLs. Doping is larger than for pure monolayer graphene deposited on Cu$(111)$, 
which shows that the interplay between nanostructuring and substrate effects leads to important changes in the electronic properties of the MLs.
As far as NRs are concerned, we have shown that all of the H-free models interact strongly with the substrate
due to the presence of dangling-bond orbitals at the edge and of edge states, both of which hybridize with the Cu d states.
In the H-terminated case, the interaction is weaker but still significant. At N edges, 
this behaviour is due to the presence of edge states. 
Interestingly, the interaction is more pronounced at the B edges, owing to a complex chemical bonding mechanism, which involves the formation
of strong B-Cu bonds.
Irrespective of the interaction at the edge, the models containing BN and C domains also interact
with Cu$(111)$ at the B-C interfaces, in full analogy with the corresponding MLs.
Some of the investigated models of NRs and MLs possess edge states and/or BN-C interface states located near the Fermi energy,
which are spin-polarized in the free-standing case. 
Here we have shown that none of the models displays significant edge/interface magnetism upon adsorption onto the Cu substrate.
However, it is plausible that some of these models may remain spin-polarized when deposited on Au$(111)$,
as occurs for passivated graphene NRs~\cite{Yan2}. 
In conclusion, our findings indicate that the interaction with the Cu substrate, in combination with nanostructuring, 
affects the structure and doping level of hybrid graphene/boron nitride systems in a non-trivial way. 
By changing the substrate type and/or by varying the size of graphene and BN domains in a controlled fashion, 
it should be possible to tune the doping level and the band gap 
of these heterostructures, with beneficial effects on their transistor properties.

\section{Acknowledgments}

We acknowledge the computational resources by the RWTH Rechenzentrum.

\end{document}